\documentclass[letterpaper, 10 pt, conference]{ieeeconf}  

\IEEEoverridecommandlockouts                              
\usepackage{amsmath}
\usepackage{multirow}
\usepackage{multicol}
\usepackage{amsfonts}
\usepackage{amssymb}
\usepackage{booktabs}
\usepackage{algorithm}
\usepackage{hyperref}

\usepackage{amsthm}
\usepackage{graphicx}
\usepackage{algpseudocode}
\algrenewcommand\algorithmicrequire{\textbf{Input:}}
\algrenewcommand\algorithmicensure{\textbf{Output:}}
\newtheoremstyle{boldtheorem}
{3pt} 
{3pt} 
{} 
{} 
{\bfseries} 
{.} 
{.5em} 
{\thmname{#1}\thmnumber{ #2}.\thmnote{ {({\normalfont\itshape#3})}}} 
\theoremstyle{boldtheorem}

\newtheorem{definition}{Definition}
\newtheorem{assumption}{Assumption}

\newtheorem{proposition}{Proposition}
\newtheorem{corollary}{Corollary}
\newtheorem{theorem}{Theorem}

\allowdisplaybreaks

\title{\LARGE \bf
L2O-CCG: Adversarial Learning with Set Generalization for\\
Adaptive Robust Optimization
}

\author{Zhiyi Zhou, J\'an Drgo\v na, Yury Dvorkin
\thanks{This work was supported in part by the
US Department of Energy (DOE) Advanced Research Projects Agency–Energy under
Grant DEAR000130010.13039/100000001, by National Science
Foundation under Grant OISE 2330450, and by the US DOE, Office of Science, ASCR program under the Scientific Discovery through Advanced Computing (SciDAC) Institute “LEADS: LEarning-Accelerated Domain Science”.}
\thanks{Zhiyi Zhou, J\'an Drgo\v na, and Yury Dvorkin are with the Johns Hopkins University, Baltimore MD, U.S. ({zzhou124, jdrgona1, ydvorki1}@jh.edu).}%
}

\begin{document}

\maketitle
\thispagestyle{empty}
\pagestyle{empty}

\begin{abstract}

The adversarial subproblem in two-stage adaptive robust optimization (ARO), which identifies the worst-case uncertainty realization, is a major computational bottleneck. This difficulty is exacerbated when the recourse value function is non-concave and the uncertainty set shifts across applications. Existing approaches typically exploit specific structural assumptions on the value function or the uncertainty set geometry to reformulate this subproblem, but degrade when these assumptions are violated or the geometry changes at deployment. To address this challenge, we propose L2O-CCG, a bi-level framework that enables the integration of structure-aware adversarial solvers within the constraint-and-column generation (CCG) algorithm. As one instantiation, we develop a generalizable adversarial learning method, which replaces solver-based adversarial search with a learned proximal gradient optimizer that can generalize across  uncertainty set geometries without retraining. Here, an inner-level neural network approximates the recourse value function from offline data, while an outer-level pre-trained mapping generates iteration-dependent step sizes for a proximal gradient scheme. We also establish out-of-distribution convergence bounds under uncertainty set parameter shifts, showing how the trajectory deviation of the learned optimizer is bounded by the uncertainty set shift. We illustrate performance of the L2O-CCG method on a building HVAC  management task. \footnote{Codes are available in: \url{https://github.com/Zhiyi-miyo/L2O-CCG}}

\end{abstract}

\section{Introduction}

\begin{table*}[!t]
\centering
\caption{Comparison of uncertainty sets}
\label{tab:uncertainty_sets}
\begin{tabular}{p{4.2cm} p{4.0cm} p{4.0cm} p{4.0cm}}
\hline
\textbf{Property} 
& \textbf{Polyhedral} 
& \textbf{Smooth (Ellipsoid)} 
& \textbf{Data-driven (W-bell)} \\
\hline

Typical form 
& $\{\xi: A\xi \le b,\ \|\xi\|_1 \le \Gamma\}$ 
& $\{(\xi-\xi^0)^\top \Sigma^{-1} (\xi-\xi^0) \le 1\}$ 
& Learned or implicit sets (VAE, GAN, samples) \\

Geometric structure 
& Polytope, axis-aligned, sparse extreme points 
& Convex, smooth boundary 
& Implicit, possibly nonconvex \\

Tractability in classical RO 
& Exact reformulation (CCG, Benders) 
& Tractable under strong assumptions 
& Often intractable \\

Typical power system applications 
& Load and renewable forecast errors in OPF; contingency modeling; reserve sizing 
& Correlated renewable generation uncertainty; chance-constrained OPF; frequency regulation 
& Line rating uncertainty; data-center loads; price-responsive demand; ramping events \\

\hline
\end{tabular}
\end{table*}

Two-stage adaptive robust optimization (ARO) is a framework for sequential decision-making under uncertainty. By decomposing decisions into here-and-now choices and wait-and-see recourse, ARO represents the adaptive nature of many real-world systems, including power system operations \cite{bertsimas2012adaptive}, model predictive control (MPC) \cite{bemporad2007robust}, supply chain design \cite{bental2004adjustable}, and capital budgeting \cite{hanasusanto2015kadapt}. The resulting min-max-min optimization, however, leads to NP-hardness in practical cases,  e.g., \cite{zeng2013solving}. For example, even when the inner-most recourse problem is convex, the adversarial max-min subproblem  can be NP-hard when the inner-most value function is non-concave with respect to the uncertainty vector \cite{zeng2013solving}. Classical methods rely on decomposition algorithms such as column-and-constraint generation (CCG) \cite{zeng2013solving}, which iteratively solve adversarial subproblems to identify worst-case scenarios. In large-scale applications, these adversarial solves dominate the computational cost, motivating methods which can accelerate the verifiable worst-case search.

Recent machine learning approaches aim to reduce this computational burden in ARO. Dumouchell et al \cite{dumouchelle2024neur2ro} train a neural network (NN) to approximate the inner-most value function and embed it into the max-min subproblem as an MILP via ReLU reformulation \cite{fischetti2018deep,grimstad2019relu}. While this approach avoids repeatedly solving the inner recourse optimization problem, the resulting adversarial problem remains a MILP whose size grows with the network and must be re-solved at every CCG iteration. Bertsimas and Kim \cite{bertsimas2024machine} bypass iterative adversarial search by solving many ARO instances offline and learning a direct mapping from problem parameters to optimal solutions. This strategy removes online adversarial solves but requires massive offline computation, and its solution quality degrades when the tested uncertainty set differs from the training. Brenner et al \cite{brenner2024deep} uses variational autoencoders to generate realistic adversarial scenarios through projected gradient ascent in a latent space, reducing conservatism in high-dimensional settings. But they assume specific uncertainty sets and may require retraining when the uncertainty set changes. These limitations highlight the need for learning-based methods that both scale with problem size and remain robust under uncertainty set shifts. 

In parallel, the learning-to-optimize (L2O) literature has developed parameterized optimizers that learn update rules from data rather than the exact estimation value. Broadly, there exsit two categories: supervised approaches \cite{sambharya2024learning, King2024} learn from datasets of solved instances to warm-start or accelerate classical solvers; and self-supervised approaches \cite{andrychowicz2016learning,li2017learning,tang2025} train parameterized update rules via algorithm unrolling without requiring pre-solved solutions. Recent work  \cite{tang2025,li2017learning} have established convergence guarantees for the self-supervised one. Martin and Furieri \cite{martin2024learningconvergence} develop convergent-by-design architectures. By learning parameterized update rules rather than repeatedly solving optimization subproblems, these approaches can alleviate the computational burden while reducing reliance on large offline datasets of solved instances. Building on this direction, \cite{kotary2025} proposes an L2O framework for solving parametric constrained bi-level problems; however, the method does not provide theoretical guarantees on convergence or robustness under out-of-distribution (OOD) instances.

In this paper, we introduce L2O-CCG, a learning-to-optimize framework that accelerates the adversarial search within CCG while generalizing across unseen uncertainty sets without retraining. The main contributions are:
\begin{itemize}
    \item \textbf{A bi-level learning framework for ARO}: We propose a modular architecture where the inner level approximates the recourse value function with a trainable surrogate, and the outer level solves the adversarial subproblem via a configurable optimizer that can be adapted to the structural properties of the surrogate and the uncertainty set. As a concrete instantiation, we pair a set-encoder NN with an LSTM-based learned optimizer.
    \item \textbf{Set-agnostic generalization via learned operators}: We design the learned optimizer so that uncertainty set parameters enter only through analytical proximal operators, not the learned parameters. This decoupling enables generalization across box, polyhedral, ellipsoidal, and GMM uncertainty sets without retraining.
    \item \textbf{Out-of-distribution convergence guarantees}: We establish convergence bounds showing that the trajectory deviation of the learned optimizer under set parameter shifts is controlled by the proximal operator discrepancy between training and test-time sets, independent of the learned parameters.
\end{itemize}
By testing on a HVAC management problem, we demonstrate that L2O-CCG achieves $10$--$100\times$ speedup over classical methods with comparable solution quality across four uncertainty geometries.

\section{Preliminaries}
Consider the general form of a two-stage ARO problem:
\begin{equation}
\label{eq:tsro}
    \min_{x \in \mathcal{X}} \; h_1(x) + \max_{\xi \in \mathcal{U}(\Gamma,\omega)} Q(x,\xi),
\end{equation}
where the inner-most (recourse) value function is defined as
\begin{equation}
\label{eq:recourse}
    Q(x,\xi) = \min_{y \in \mathcal{Y}(x,\xi)} h_2(y)
\end{equation}
and where $x$ denotes first-stage (here-and-now) decisions made prior to the realization of uncertainty, $y$ represents second-stage (recourse) decisions that adapt after observing uncertainty realization $\xi$. The uncertainty vector $\xi$ belongs to a prescribed uncertainty set $\mathcal{U}(\Gamma, \omega)$ with budget $\Gamma$ and shape parameter $\omega$ that characterizes the geometry of the set (e.g., covariance in ellipsoidal sets or component-wise bounds in box sets). The feasible recourse region is given by
\begin{equation}
\mathcal{Y}(x,\xi) = \{y \mid g_1(y,x,\xi)\le 0,\; g_2(y,x,\xi)=0 \}.
\end{equation}

Classical approaches typically rely on decomposition methods to solve \eqref{eq:tsro}, including Benders-type and cutting-plane decompositions \cite{bertsimas2012adaptive}. Among these, CCG \cite{zeng2013solving} has emerged as one of the most effective methods in practice, which iterates between a master problem and a worst-case subproblem. At iteration $k$, the master problem
\begin{align} 
    \min_{x,\theta} & \; h_1(x) + \theta \label{eq:master}\\
    \text{s.t. } &\theta \ge Q(x,\xi^i), i = 1,\ldots,k, \quad x \in \mathcal{X}, \nonumber
\end{align}
is solved over accumulated uncertainty scenarios $\{\xi^i\}_{i=1}^k$. Given the master solution $x^k$, the adversarial subproblem
\begin{equation} \label{eq:subproblem}
    \xi^{k+1} \in \arg\max_{\xi \in \mathcal{U}} Q(x,\xi).
\end{equation}
identifies the worst-case uncertainty. If no scenario increases the objective beyond tolerance $\epsilon$, the algorithm terminates. Otherwise, $\xi^{k+1}$ is added and the procedure repeats. The CCG performance and convergence  depend on the tractability of \eqref{eq:subproblem}, which in turn is impacted by the structure of $Q(x,\xi)$ and the geometry of $\mathcal{U}$.

\textbf{Structure of $Q(x,\xi)$.} In general, $Q(x,\xi)$ is neither convex nor concave in $\xi$. This occurs  when uncertainty affects system dynamics or constraint coefficients, which is common in control, energy, and resource allocation problems. In such cases, $\max_{\xi \in \mathcal{U}} Q(x,\xi)$ may contain multiple local maxima, and the adversarial subproblem becomes NP-hard even when the inner recourse problem is convex. Table~\ref{tab:Q_structure} summarizes typical $Q$-structures arising in common models.
    
\textbf{Geometry of $\mathcal{U}$.} The uncertainty set determines the feasibility  region for~\eqref{eq:subproblem}. Polyhedral sets allow for vertex enumeration and are compatible with MILP-based CCG. Ellipsoidal sets capture correlations but require quadratic constraints in the adversarial problem, making each CCG iteration substantially more expensive than for linear uncertainty sets. Data-driven sets (e.g., GMM density level sets) can represent multi-modal uncertainty but are non-convex, introducing binary variables or big-M disjunctions. Table~\ref{tab:uncertainty_sets} summarizes common uncertainty set geometries.
    
\textbf{Joint effect.} Table~\ref{tab:method_selection} shows that when $Q$ is concave in $\xi$, classical methods solve~\eqref{eq:subproblem} efficiently regardless of set geometry. However, for non-concave $Q$, solver-based reformulations become intractable or lose global optimality guarantees, regardless of  $\mathcal{U}$. Moreover, methods tailored to a specific uncertainty geometry degrade under distributional shift~\cite{bertsimas2024machine, brenner2024deep}. These limitations motivate L2O-based approaches that approximate the worst-case search, without relying on specific assumptions for $Q$ or $\mathcal{U}$ structures.

\begin{table}[tb]
\centering
\caption{Comparison of Q structure for recourse problems}
\label{tab:Q_structure}
\begin{tabular}{p{4.3cm} p{3.3cm}}
\hline
\textbf{Inner Problem} 
& \textbf{Q structure} \\
\hline

LP with $\xi$ in RHS
& piecewise affine and convex \\

QP with $\xi$ in RHS
& piecewise affine and convex \\

QCQP with $\xi$ in RHS
& piecewise affine and convex \\

QP with $\xi$ in linear coefficients
& nonconvex and nonconcave \\

QCQP with $\xi$ in quadratic coefficients
& nonconvex and nonconcave \\
\hline
\end{tabular}
\end{table}

\begin{table}[!tb]
\centering
\caption{Optimizer selection for $\max_{\xi \in \mathcal{U}} Q(x,\xi)$ under different uncertainty geometries and $Q$-structures.}
\label{tab:method_selection}
\renewcommand{\arraystretch}{1.15}
\setlength{\tabcolsep}{3pt}
\begin{tabular}{p{0.8cm} p{3.0cm} p{3.6cm}}
\hline
\textbf{Set $\mathcal{U}$} 
& \textbf{$Q$-structure} 
& \textbf{Optimizer} \\
\hline
\multirow{5}{*}{\parbox{1.4cm}{Poly-\\hedral}}
& Concave, affine 
& LP~\cite{bertsimas2004price}; exact \\
& Concave, quadratic 
& QP~\cite{bental1999robust}; exact \\
& Concave, ICNN 
& Gradient-based: projected ascent / Frank--Wolfe~\cite{amos2017input}; convergent \\
& Concave non-smooth, ReLU 
& MILP reformulation~\cite{anderson2020strong}; exact but expensive \\
& Non-concave, general NN 
& L2O / multi-start~\cite{andrychowicz2016learning}; heuristic \\
\hline
\multirow{5}{*}{\parbox{1.4cm}{Ellip-\\soidal}}
& Concave, affine 
& Closed-form~\cite{bental2009robust}; exact \\
& Concave, quadratic 
& QCQP~\cite{conn2000trust}; exact \\
& Concave, ICNN 
& Projected ascent~\cite{boyd2004convex}; efficient \\
& Concave non-smooth, ReLU 
& MILP / QCQP reformulation~\cite{anderson2020strong}; scales poorly \\
& Non-concave, general NN 
& L2O / gradient~\cite{andrychowicz2016learning}; heuristic \\
\hline
\multirow{5}{*}{\parbox{1.4cm}{Data-\\driven}}
& Concave, affine 
& Cutting-plane~\cite{nemirovski2004prox}; reliable \\
& Concave, quadratic 
& Frank--Wolfe; moderate \\
& Concave, ICNN 
& Gradient-based with learned projection~\cite{amos2017optnet}; scalable \\
& Concave non-smooth, ReLU 
& Oracle-based search / Learning optimizer~\cite{esfahani2018data}; approximate \\
& Non-concave, general NN 
& L2O~\cite{chen2017learning}; scalable, heuristic \\
\hline
\end{tabular}
\end{table}


\section{Proposed L2O-CCG Framework}\label{sec:framework}
\subsection{Bi-Level Learning Structure}\label{sec:bilevel}

\begin{figure}[t]
    \centering
    \includegraphics[width=\columnwidth]{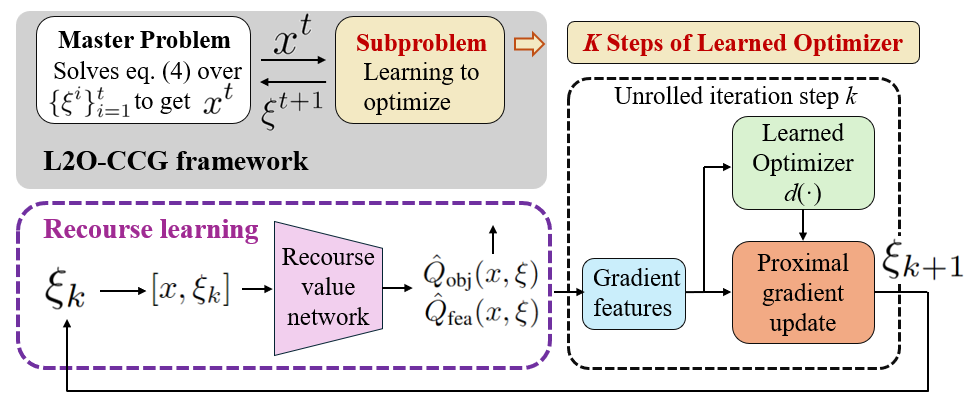}
    \caption{Diagram for the proposed L2O-CCG}
    \label{fig:dia}
\end{figure}

We propose a generalizable bi-level learning framework that replaces the computationally expensive adversarial subproblem of CCG with a learned update rule, as shown in Fig.~\ref{fig:dia}. Inner-level NNs $\hat{Q}_\text{obj}(x,\xi)$ and $\hat{Q}_\text{fea}(x,\xi)$  approximate the recourse value function and feasibility, respectively,  from offline data, while an outer-level learned optimizer $d(\cdot)$ solves the adversarial subproblem via gradient-based search based on the learned $\hat{Q}_\text{obj}(x,\xi)$ and $\hat{Q}_\text{fea}(x,\xi)$ values.

At each iteration $t$, the modified procedure is:
\begin{enumerate}
    \item \textbf{Master problem}: Solve \eqref{eq:master} to obtain $x^t$.
    \item \textbf{Learned adversarial}: Starting from multiple initial points $\{\xi_0^{(m)}\}_{m=1}^M$ sampled from $\mathcal{U}$, run $K$ steps of the learned optimizer:
    \begin{equation}\label{eq:l2o_update}
        \xi_{k+1} = \xi_k - d(z_k),
    \end{equation}
    where $d(z_k)$ is the learned optimizer that generates iteration-dependent step sizes from gradient features $z_k$, and the proximal operator enforces $\xi \in \mathcal{U}$. Return $\xi^{t+1} \leftarrow \arg\min_m F(\xi_K^{(m)})$. The feature vector $z_k$ includes the gradient of the value function $\hat{Q}_\text{obj}(x,\xi)$ and $\hat{Q}_\text{fea}(x,\xi)$ with respect to $\xi$. The full design is detailed in Section \ref{sec:l2o_adv}. 
    \item \textbf{Exact verification}: Evaluate the true recourse $Q(x^t, \xi^{t+1})$ by solving \eqref{eq:recourse}.
    \item \textbf{Convergence check}: If the true gap $(Q(x^t,\xi^{t+1}) - \theta^t)/(|\theta^t|+\epsilon) \le \epsilon$, terminate. Otherwise add $\xi^{t+1}$ to the master problem and repeat.
\end{enumerate}

Algorithm~\ref{alg:l2o_ccg} summarizes the complete procedure. Since the master problem and verification use exact solvers, every scenario added is valid. If the learned adversarial misses the global worst case, CCG continues to iterate and terminates finitely under standard assumptions, e.g., Zeng et al \cite{zeng2013solving} point out that polyhedral uncertainty set should contain finitely many extreme points and put into a convex recourse problem.  L2O  replaces the MILP-based solution in \cite{dumouchelle2023deep} with $M \times K$ forward passes, typically $2$--$3$ orders of magnitude faster per CCG iteration. 

The two learning modules are trained sequentially. First, the value network $\hat{Q}(\xi;\phi)$ is trained from a dataset of solved recourse problems. Then, the adversarial optimizer $d(\cdot)$ is trained for recourse value function $\hat{Q}$ over $\mathcal{U}$.

\begin{algorithm}[t]
\caption{L2O-CCG: CCG with Learned Adversarial}
\label{alg:l2o_ccg}
\begin{algorithmic}[1]
\Require Uncertainty set $\mathcal{U}$, tolerance $\epsilon$, multi-starts $M$, L2O steps $K$
\Ensure First-stage decision $x^*$, robust objective $\theta^*$

\State \textbf{Offline:}
\State Train value network $\hat{Q}(\cdot;\phi)=\big[\hat{Q}_{\text{obj}},\hat{Q}_{\text{fea}}\big]$ from data 
\State Train learned optimizer $d(\cdot)$ via~\eqref{eq:l2o_loss} 

\State \textbf{Online:}
\State $\Xi \gets \{\mathbf{0}\}$
\For{$t = 1,2,\ldots,T$}
    \State $(x^t,\theta^t) \gets$ solve master~\eqref{eq:master} over $\Xi$
    \For{$m = 1,\ldots,M$} \Comment{multi-start L2O}
        \State Sample $\xi^{(m)}_{0}$ from $\mathcal{U}$
        \For{$k = 0,1,\ldots,K-1$}
            \State $\xi^{(m)}_{k+1} \gets$ update via~\eqref{eq:l2o_update} using $d(z^{(m)}_k)$
        \EndFor
    \EndFor
    \State $\xi^{t+1} \gets \arg\min_{m\in\{1,\ldots,M\}} F\!\left(\xi^{(m)}_{K}\right)$
    \State Compute true $Q(x^t,\xi^{t+1})$ via~\eqref{eq:recourse} \Comment{verification}
    \If{$\text{gap} \le \epsilon$}
        \State \Return $(x^t,\theta^t)$
    \EndIf
    \State $\Xi \gets \Xi \cup \{\xi^{t+1}\}$
\EndFor
\end{algorithmic}
\end{algorithm}

\subsection{Adversarial Learning}\label{sec:l2o_adv}
\subsubsection{Inner-Level Surrogate}
The inner level of L2O-CCG requires a differentiable approximation of $Q(x,\xi)$. The choice of surrogate architecture should reflect the known or expected $Q$-structure, e.g., according to Table. \ref{tab:method_selection}, input-convex NN when $Q$ is known to be concave, kernel-based models for low-dimensional settings, or general feedforward networks when $Q$ is non-concave. In this work, we adopt a general NN to handle the non-concave case without structural assumptions.

We first train a NN $\hat{Q}(x,\xi;\phi)$ to approximate the recourse value function from a dataset of solved second-stage problems $ \{(x^{(i)}, \xi^{(i)}, Q^{(i)}_\text{obj}, Q^{(i)}_\text{fea})\}_{i=1}^N$. Motivated by \cite{dumouchelle2024neur2ro}, the architecture uses a set-encoder with element-wise MLPs and sum-pooling aggregation:
\begin{subequations}\label{eq:set_encoder}
\begin{gather}
    \Phi_x(x) = \sigma_{\text{post}}\!\left(\textstyle\sum_{j=1}^{n_x} \sigma_{\text{elem}}(x_j)\right)\\
    \Phi_\xi(\xi) = \sigma_{\text{post}}\!\left(\textstyle\sum_{j=1}^{n_\xi} \sigma_{\text{elem}}(\xi_j)\right),
\end{gather}
\end{subequations}
and the concatenated embedding is passed through a dual-head value network:
\begin{equation}\label{eq:dual_head}
    [\hat{Q}_\text{obj}, \; \hat{Q}_\text{fea}] = \text{MLP}_\text{val}\big([\Phi_x(x); \Phi_\xi(\xi)]\big),
\end{equation}
where $\hat{Q}_\text{obj}$ and $\hat{Q}_\text{fea}$ predict the recourse cost and constraint violation. Both heads are trained jointly with mean-squared-error loss on min-max normalized targets. The dual-head design enables the adversarial optimizer to seek scenarios that maximize cost while maintaining a  feasible recourse.

\subsubsection{Outer-Level Optimizer}
Given the trained value  $\hat{Q}(\cdot;\phi)$, the adversarial subproblem becomes $\min_{\xi \in \mathcal{U}} F(\xi)$ with the augmented objective
\begin{equation}\label{eq:augmented_obj}
    F(\xi) = \underbrace{-\hat{Q}_\text{obj}(\xi; x)}_{f(\xi)} + \underbrace{\alpha \, d(\xi; \mathcal{U}) + \beta \max\!\big(0, \hat{Q}_\text{fea}(\xi; x)\big)}_{r(\xi)},
\end{equation}
decomposed into a smooth component $f$ (differentiable through the NN) and a non-smooth component $r$ handled by the proximal operator. 

The outer-level optimizer can likewise be chosen based on the surrogate structure, e.g, according to Table. \ref{tab:method_selection}, projected gradient ascent for concave surrogates, Frank--Wolfe for polyhedral sets with linear surrogates, or MILP reformulation for ReLU networks. For the non-concave general-NN setting targeted here, we adopt a self-supervised L2O approach based on \cite{song2024towards}, where the optimizer is trained via algorithm unrolling without requiring pre-solved adversarial instances.

Specifically, the update rule~\eqref{eq:l2o_update} expands as:
\begin{subequations}\label{eq:l2o_detail}
\begin{align}
    v_k &= Q_k \, v_{k-1} + (I - Q_k) \nabla f(\xi_k), \label{eq:momentum}\\
    \bar{\xi}_{k+1} &= \xi_k - R_k \nabla f(\xi_k) - B_k \, v_k, \label{eq:gradient_step}\\
    \xi_{k+1} &= \mathrm{prox}_{r, R_k}(\bar{\xi}_{k+1}), \label{eq:prox_step}
\end{align}
\end{subequations}
where $R_k, Q_k, B_k \in (0,1)^{n_\xi}$ are diagonal step-size and momentum matrices generated by $d(z_k)$, and $v_k$ is a momentum variable. The proximal operator
\begin{equation}\label{eq:prox_def}
    \mathrm{prox}_{r, R_k}(\bar{\xi}) = \arg\min_{\xi} \left\{ r(\xi) + \tfrac{1}{2} \|\xi - \bar{\xi}\|^2_{R_k^{-1}} \right\}
\end{equation}
enforces set membership while accounting for the non-smooth penalty in $r$. The learned optimizer $d(\cdot)$ is parameterized as a mapping $\phi$:
\begin{equation}\label{eq:lstm}
    (R_k, Q_k, B_k),\; s_k = \phi_\psi(z_k, s_{k-1}),
\end{equation}
where $s_{k-1}$ is the hidden state and the feature vector is
\begin{equation}\label{eq:features}
    z_k = [\nabla f(\xi_k), \; g_k, \; v_{k-1}],
\end{equation}
consisting of the current gradient $\nabla f(\xi_k)$,  subgradient $g_k \in \partial r(\xi_k)$, and the previous momentum $v_{k-1}$. Outputs are passed through sigmoid activations to ensure bounded step sizes, as required by the convergence theory in~\cite{song2024towards}.

We opt for $z_k$ to include only gradient-based features, that is the uncertainty set parameters $(\Gamma, \omega)$ are excluded from the $\phi_\psi(\cdot)$ input. Instead, these set parameters enter exclusively through the proximal operator~\eqref{eq:prox_step}. This decoupling yields a geometry-agnostic step-size policy: the mapping $\phi_\psi(\cdot)$ generalizes across uncertainty sets  and parameters without retraining, while the proximal operator handles set-specific geometry. We provide formal OOD guarantees  in   Section \ref{Sec:guarantees}.

The proximal step~\eqref{eq:prox_step} is the only component that depends on the uncertainty set geometry. We derive closed-form or efficient proximal operators for common set geometries:
\begin{itemize}
    \item \textbf{Box} $\mathcal{U}_\text{box}$: Clamping to $[-\theta_j, \theta_j]$ followed by $\ell_1$-ball projection for budget $\|\xi\|_1 \le \Gamma$.
    \item \textbf{Polyhedral} $\mathcal{U}_\text{poly}$: Box projection followed by iterative halfplane projection (Dykstra-style) for $H\xi \le h$.
    \item \textbf{Ellipsoidal} $\mathcal{U}_\text{ellip}$: Scaling projection $\xi \leftarrow \xi \cdot \min(1, \Gamma / \|\xi\|_{\Sigma^{-1}})$.
    \item \textbf{GMM level set} $\mathcal{U}_\text{gmm}$: Projection onto the nearest component ellipsoid via Mahalanobis distance.
\end{itemize}
All operators are differentiable almost everywhere, enabling end-to-end training of $\phi_\psi(\cdot)$ through backpropagation. The parameters $\psi$ are trained by unrolling $K$ adversarial steps and minimizing the cumulative augmented objective:
\begin{equation}\label{eq:l2o_loss}
    \min_\psi \; \mathbb{E}_{x \sim \mathcal{D}_x} \left[ \sum_{k=1}^{K} F(\xi_k; x, \mathcal{U}) \right],
\end{equation}
where the trajectory $\{\xi_k\}_{k=0}^K$ is generated by the update rule~\eqref{eq:l2o_detail} with parameters from~\eqref{eq:lstm}. Gradients are propagated via truncated backpropagation through time. At inference, the trained optimizer runs from $M$ random initial points sampled from $\mathcal{U}$, and the best solution across restarts is returned as the adversarial scenario.

\section{Formal OOD Guarantees for\\ Uncertainty Sets} 
\label{Sec:guarantees}
This section establishes theoretical guarantees for the proposed L2O-CCG when the uncertainty-set parameters differ from those used during training. In particular, we analyze how changes in the uncertainty-set parameters $(\Gamma,\omega)$ affect the convergence behavior of the learned optimizer.

Our analysis proceeds in three steps. First, we establish convergence of the learned optimizer in the in-distribution (InD) setting where the uncertainty-set parameters belong to the training domain. Second, we analyze how deviations in the uncertainty-set parameters induce feature shifts in the learned optimizer and quantify their impact on the optimization trajectory. Finally, we derive bounds on the resulting performance degradation and show that the analysis applies to a broad class of uncertainty sets through a unified penalty representation.

Let $F(\xi) = f(\xi) + r(\xi)$ where $f(\xi) = - \hat{Q}_\text{obj}(\xi;x)$ represents the piecewise-smooth objective component, while $r(\xi) = \alpha d(\xi;\Gamma,\theta) + \beta \max(0, \hat{Q}_\text{fea}(\xi,x))$ represents the non-smooth penalty components.

\begin{assumption}[Piecewise $L$-smoothness]\label{ass:L-smooth}
    The learned inner recourse  $f(\xi) = \hat{Q}_\text{obj}(x,\xi)$ is a piecewise $L$-smooth function. There exists a finite partition $\{\mathcal{P}_1, ..., \mathcal{P}_M\}$ of the domain such that on each piece $\mathcal{P}_i$, there exists $L_i \in \mathbb{R}$:
    \begin{equation}
        ||\nabla f(\xi_1) - \nabla f(\xi_2)|| \leq L_i ||\xi_1 - \xi_2||, \forall\xi_1,\xi_2 \in \mathcal{P}_i.
    \end{equation}
    Let $L \triangleq \max_i L_i $ be the maximum local Lipschitz constant.
    \hfill \qed
\end{assumption}

\begin{definition}[InD Domain and OOD Domain]\label{def:InD_domain}
    Let $\mathcal{S}_P \subseteq \mathbb{R}^+ \times \Omega$ be a compact set of uncertainty parameters used during training. Then the in-distribution (InD) L2O problem corresponds to $(\Gamma_\text{in},\omega_\text{in}) \in \mathcal{S}$. The OOD domain is $\mathcal{S}_O \triangleq \mathbb{R}^+ \times \Omega \backslash \mathcal{S}$. OOD scenarios are $(\Gamma_\text{out},\omega_\text{out})\in \mathcal{S}_O$.
    \hfill \qed
\end{definition}

We first establish convergence of L2O-CCG in the InD setting, where the configurations  coincide during training and testing. Proposition~\ref{pro:InD} shows that under the piecewise $L$-smoothness assumption, the learned update rule achieves the standard proximal-gradient descent guarantee up to additional terms that appear when the trajectory crosses piecewise boundaries of the value function.

\begin{proposition}[InD Convergence with Boundary Transition \cite{song2024towards}]  \label{pro:InD}
    For the composite $F(\xi) = f(\xi) + r(\xi)$ with piecewise $L$-smooth $f(\cdot)$, the optimal L2O configuration is:
    \begin{equation}
        N_1 = \frac{1}{L} \mathbf{1}, \quad N_2 = 0
    \end{equation}
    Under these optimal settings, when crossing from piece $\mathcal{P}_i$ to $\mathcal{P}_j$ at boundary $\xi_b$, define gradient jump as $\Delta \xi_b \triangleq ||\nabla f_j(\xi_b) - \nabla f_i(\xi_b)||$. The per-iteration bound is given as:
    \begin{equation}
        F(\xi_k)- F(\xi_{k-1}) \leq -\frac{1}{2L}||\xi_k - \xi_{k-1}||^2 + \frac{1}{2L} \Delta \xi,
    \end{equation}
    which can be accumulated across multiple steps:
    \begin{equation}
        F(\xi_K) - F(\xi^*) \leq \frac{L}{2K} ||\xi_0-\xi^*||^2 + \frac{1}{2LK}\sum_k \Delta \xi_k
    \end{equation}
    where $\Delta \xi_k = 0$ is no crossing.
\end{proposition}

Then motivated by \cite{song2024towards}, we consider two trajectories produced by the same learned update rule $d(\cdot)$: InD trajectory $\{\xi_{k,\text{in}}, z_{k,\text{in}}\}_{k\ge 0}$ generated using
$(\Gamma_\text{in},\omega_\text{in})$, and OOD trajectory $\{\xi_{k,\text{out}}, z_{k,\text{out}}\}_{k\ge 0}$ generated using $(\Gamma_\text{out},\omega_\text{out})$.

Here we apply gradient-only features: 
\begin{equation}
    z = [\nabla f(\xi)^\top, g^\top]^\top, g\in \partial r(\xi;\Gamma,\omega)
\end{equation}

\begin{definition}[Virtual Feature] \label{def:virtual_feat}
    Consider an OOD uncertainty variable $\xi_\text{out} \in \mathcal{U}(\Gamma_\text{out},\omega_\text{out})$ with parameters $(\Gamma_\text{out},\omega_\text{out})\notin \mathcal{S}$ and a corresponding InD variable $\xi_\text{in} \in \mathcal{U}(\Gamma_\text{in},\omega_\text{in})$ with $(\Gamma_\text{in},\omega_\text{in}) \in \mathcal{S}$. Define the virtual feature as the difference between input features:
    \begin{equation}
        \delta = z_\text{out} - z_\text{in} = [(\nabla f(\xi_\text{out})-\nabla f(\xi_\text{in}))^\top, (g_\text{out}-g_\text{in})^\top]^\top
    \end{equation}
    where the first component is the gradient difference due to different iterations as $f(\cdot)$ doesn't depend on $\Gamma$ and $\omega$, and the second component refers to the subgradient difference where $g_\text{in} \in \partial r(\xi_\text{in};\Gamma_\text{in},\omega_\text{in}), g_\text{out} \in \partial r(\xi_\text{out};\Gamma_\text{out},\omega_\text{out})$.
    \hfill \qed
\end{definition}

Building upon the  InD guarantee  in Proposition \ref{pro:InD}, we now consider OOD scenarios where only the uncertainty set parameters $\Gamma,\omega$ change while the piecewise smooth recourse component $f(\xi)$ remains identical. This constraint-only shift simplifies the virtual-feature structure and makes the deterioration explicitly traceable to boundary violations.

The OOD deterioration can be quantified through feature shift $\delta_k$, and then transferred to the iterate shift $s_k$ via a virtual Jacobian. Similar to \cite{song2024towards}, the difference between two updates $d(z_{k,\text{out}})$ and $d(z_{k,\text{in}})$ is represented by a linear map applied to the feature shift. 

We assume that the learned update rule $d(\cdot)$ is differentiable almost everywhere and Lipschitz on the feature domain of interest. In particular, for each iteration $k$, there exists a virtual Jacobian matrix $J_{k}\in\mathbb{R}^{d\times p}$ such that
\begin{align}
d(z_{k,\text{out}}) &= d(z_{k,\text{in}}) + J_{k}(z_{k,\text{out}} - z_{k,\text{in}}) \nonumber\\
&= d(z_{k,\text{in}}) + J_{k} \delta_k,
\label{eq:virtual_jacobian}
\end{align}
and $\|J_{k}\| \le C_J \sqrt{n}$ for some constant $C_J>0$, where $n$ is the dimension of $\xi$.

Under this assumption, the variable shift satisfies
\begin{subequations}
    \begin{gather}
        s_{k+1}= s_k + J_{k} \delta_k\\
        \|s_{k+1}\| \le \|s_k\| + C_J \|\delta_k\| \label{eq:s_recursion}
    \end{gather}
\end{subequations}
which shows that the key is to quantify $\|\delta_k\|$,
which in our setting is induced by $(\Gamma,\omega)$ shift through the penalty term.

\begin{theorem}[Per-Iteration OOD Descent Bound]\label{the:OOD_itera}
    Under the piecewise $L$-smoothness assumption and InD-optimal parameters ($N_1 = \frac{1}{L} \mathbf{1}, \quad N_2 = 0$), gradient-only features and the virtual-feature construction, for any OOD pair $(\Gamma_\text{out},\omega_\text{out}) \in \mathcal{P}_O$, the per-iteration ascent satisfies:
    \begin{align}
        &F(\xi_k+s_k) - F(\xi_{k-1} + s_{k-1}) \nonumber\\
        \leq& -\frac{||\nabla f(\xi_{k-1}+s_{k-1})+g_{k,\text{out}}||}{2L} \nonumber\\
        & + L ||\text{diag}(J_{1,k-1}\delta_k)(\nabla f+g_\text{out})||^2  \\
        &+ L ||\frac{\nabla f(\xi_{k-1}+s_{k-1})+g_{k,\text{out}}-\nabla f(\xi_{k-1})-g_{k,\text{in}}}{2L} \nonumber\\
        &\quad\quad\quad- J_{2,k-1}\delta_k||^2 \nonumber
    \end{align}
    where the virtual Jacobians satisfy $||J_i||_F \leq C_i \sqrt{n}$
\end{theorem}

In Theorem \ref{the:OOD_itera}, the first term is exactly the ideal InD proximal-gradient improvement. The two extra quadratic penalties vanish identically whenever both trajectories stay strictly inside their respective budgeted regions.

The subgradient difference decomposes as:
\begin{equation}
    \Delta g_k = ||g_{k,\text{out}}- g_{k,\text{in}}||\leq \|\Delta_r^\Gamma(\xi)\| + \|\Delta_r^\omega(\xi)\|,
\end{equation}
where
\begin{align}
    \Delta_r^\Gamma(\xi)
&=
\nabla_\xi r(\xi;\Gamma_\text{out},\omega_\text{in})
-\nabla_\xi r(\xi;\Gamma_\text{in},\omega_\text{in}),
\label{eq:delta_r_gamma}
\\
\Delta_r^\omega(\xi)
&=
\nabla_\xi r(\xi;\Gamma_\text{in},\omega_\text{out})
-\nabla_\xi r(\xi;\Gamma_\text{in},\omega_\text{in}).
\label{eq:delta_r_omega}
\end{align}

Here we quantify the $\Gamma$-shift and $\omega$-shift independently. The $\Gamma$-shift affects only the budget component of the penalty. The $\omega$-shift changes both (i) the violation level and the chain-rule scaling, producing an anisotropic preconditioning mismatch.

\begin{corollary}[Subgradient Difference]
    Let $S(\xi;\omega)$ denote the smoothed budget statistic, and let the budget penalty be a smooth function of $t(\xi;\Gamma,\omega)\triangleq S(\xi;\omega)-\Gamma$. The subgradient difference is:
    \begin{equation}
        \Delta g_k = (\alpha(t_\text{out})-\alpha(t_\text{in})) \nabla_\xi S (\xi;\omega_\text{in}) + \Delta_r^\omega(\xi),
    \end{equation}
    where $\Delta_r^\omega(\xi)$ follows the Lipschitz split with factors $\frac{1}{\omega_{\min}^2}$. The OOD penalty terms in Theorem \ref{the:OOD_itera} are bounded by:
    \begin{equation}
        O(L_\alpha (\xi)^2 |\Delta \Gamma|^2 ||\nabla_\xi S||^2 + \frac{||\Delta \omega \odot (\xi-\xi^0)||^2+||\Delta \omega||^2}{\omega_{\min}^4})
    \end{equation}
\end{corollary}

The penalty is exactly zero on all iterations where neither trajectory touches its boundary. The per-iteration local bound naturally extends to the global multi-iteration rate via telescoping.

\begin{theorem}[Multi-Iteration OOD Rate]
    The $K$ iterations' convergence rate in the OOD scenarios is upper bounded by:
    \begin{align}
        &\min_{1\le k \le K} [F(\xi_k+s_k)-F(\xi^*+s_{k-1})] \nonumber\\
        \leq& \frac{L}{2} ||\xi_0 + s_0 - \xi^*-s^*||^2 \nonumber\\
        &-\frac{L}{2}||\xi_K+s_K-\xi^*-s^*||^2 \\
        &+ L\sum_{k=1}^K (\xi_k+s_k-(\nabla f + g_{k,\text{out}})/L)^\top \nonumber\\
        &\quad(\xi_k+s_k-\xi^*-s^*)  \nonumber  \end{align}
\end{theorem}

The summation term is bounded by $O(\sum_k ||\Delta g(\Delta \Gamma,\Delta \omega)||^2)$ and is identically zero on all iterations that remain interior to both budgeted regions.

The proposed OOD analysis does
not rely on any combinatorial or linear structure of the uncertainty set.
Instead, it only requires that the uncertainty set be encoded through a
soft penalty function $r(\xi;\omega)$ with parameters $\omega$. This representation can cover a broad family of uncertainty sets. Table~\ref{tab:uncertainty_generalization} summarizes several examples and shows how parameter shifts correspond to the feature shifts analyzed above. For the polyhedral set, the budget statistic $S(\xi;\omega)$ represents a weighted $\ell_1$ deviation from the nominal point $\xi^0$. A change in $\Gamma$ shifts the threshold defining the feasible region, while changes in $\omega$ rescale individual coordinates and thus alter the anisotropy of the uncertainty region. For the ellipsoidal set, the budget statistic is a quadratic form defined by the covariance matrix $\Sigma$. A change in $\Gamma$ modifies the confidence radius of the ellipsoid, while changes in $\Sigma$ alter its orientation and axis lengths, effectively introducing a quadratic preconditioning mismatch in the penalty gradient. For data-driven sets, the budget statistic is defined through a learned representation $\rho(\xi;\omega)$, such as those arising from generative or distributionally robust models. In this case, $\Gamma$ controls the ambiguity radius, while the parameters $\omega$ define the geometry of the learned uncertainty representation. 

So in all cases listed above, the penalty
takes the generic form
$r(\xi;\omega)=\phi(S(\xi;\omega)-\Gamma)$, where $S(\cdot)$ is the corresponding
budget statistic and $\phi$ is a smooth hinge-type function.
Consequently, the penalty gradient admits the unified representation
$\nabla_\xi r(\xi;\omega)=\alpha(S(\xi;\omega)-\Gamma)\nabla_\xi S(\xi;\omega)$,
with $\alpha=\phi'$.
This structure allows the OOD mismatch $\Delta_r(\xi)$ to be decomposed into
a threshold-induced $\Gamma$-shift term and a geometry-induced parameter-shift
term, both of which can be quantified using the same analysis framework.

\begin{table}[!t]
\centering
\caption{Uncertainty-set parameterization and correspondence with the proposed OOD analysis}
\label{tab:uncertainty_generalization}
\begin{tabular}{p{1.0cm} p{3.0cm} p{3.2cm}}
\hline
\textbf{Set} 
& \textbf{Budget statistic $S(\xi;\omega)$} 
& \textbf{OOD shift} \\
\hline
Polyhedral
& $\displaystyle S(\xi;\omega)=\sum_i \frac{|\xi_i-\xi_i^0|}{\omega_i}$ 
& $\Gamma$-shift: threshold shift at budget boundary; \newline
$\omega$-shift: anisotropic coordinate scaling \\[0.6em]

Ellipsoidal
& $\displaystyle S(\xi;\Sigma)=(\xi-\xi^0)^\top \Sigma^{-1}(\xi-\xi^0)$
& $\Gamma$-shift: radius/confidence level change; \newline
$\Sigma$-shift: quadratic preconditioning mismatch \\[0.6em]

Data-driven
& $\displaystyle S(\xi;\omega)=\rho(\xi;\omega)$
& $\Gamma$-shift: ambiguity radius change; \newline
$\omega$-shift: learned representation geometry shift \\
\hline
\end{tabular}
\end{table}
\section{Experiments}
All experiments are conducted on an Intel Core i7-10875H CPU and an NVIDIA GeForce RTX 2060 GPU (6 GB VRAM). The algorithms were implemented in Python 3.10 using PyTorch 2.2 \cite{paszke2019pytorch} and Gurobi 13.0 \cite{gurobi}. NNs were trained using GPU acceleration, whereas optimization subproblems were solved on CPU.

\subsection{Problem Settings}
We consider a two-stage ARO problem arising in a   HVAC thermal dynamics model of buildings:
\begin{subequations}\label{eq:robust_mpc}
    \begin{align}
        \min_{u_0} \max_{\xi \in \mathcal{U}(\Gamma,\theta)} \min_{x_t,u_t} { } & x^\top_N P_f x_N +  \sum_{t=1}^{N-1} (x_t^\top P x_t + u_t^\top R u_t)   \nonumber\\
        \text{s.t. } & x_{t+1} = A_t(\xi) x_t + B_t(\xi) u_t, \forall t\\
        & A_t(\xi) = A_{t,0} + a_t \xi , \forall t\\
        & B_t(\xi) = B_{t,0} + b_t \xi , \forall t\\
        & \underline{x} \le x_t \le \overline{x}, \underline{u} \le u_t\le\overline{u}, \forall t\\
        &\Delta \underline{u} \le u_t -u_0\le \Delta \overline{u}, \forall t \label{eq:deviation_bounds}
    \end{align}
\end{subequations}
where the temperature state dimension $n_x = 4$ in coupled building zones, input dimension $n_u = 2$, corresponding to HVAC control signals. The planning horizon is $N = 10$, and uncertainty dimension is $n_\xi = 5$. The uncertainty vector $\xi$ captures uncertainty in heat transfer coefficients between zones, ambient heat exchange, thermal damping, and HVAC actuation efficiency. Constraint~\eqref{eq:deviation_bounds} bounds the deviation of recourse actions from the first-stage decision, modeling limited recourse flexibility.

We apply a LSTM-based mapping for the adversarial learning. The optimizer runs $K = 40$--$60$ adversarial steps with $10$--$15$ random restarts at inference time. We evaluate performance across four uncertainty sets:

Box with budget: $\mathcal{U}_{\mathrm{box}} = \{ \xi : |\xi_j| \le \theta_j, \; \textstyle\sum_{j=1}^{n_\xi} |\xi_j| \le \Gamma \}.$

Polyhedral with budget: $\mathcal{U}_{\mathrm{poly}} \!=\! \{ \xi : H \xi \!\le\! h, \; |\xi_j| \!\le\! \theta_j, \; \textstyle\sum_j |\xi_j| \!\le\! \Gamma \}$ where $H \in \mathbb{R}^{m \times n_\xi}$ encodes coupling constraints between related uncertainty components (i.e., joint limits on spring stiffness perturbations).

Ellipsoid with budget: $\mathcal{U}_{\mathrm{ellip}} = \left\{ \xi : \xi^\top \Sigma^{-1} \xi \le \Gamma^2 \right\}$ with $\Sigma$ encoding correlations between uncertainties.

GMM density level set (nonconvex): $\mathcal{U}_{\mathrm{gmm}} = \left\{ \xi : \textstyle\sum_{c=1}^C w_c \, \mathcal{N}(\xi \mid \mu_c, \Sigma_c) \ge \rho \right\}$.

We compare the proposed L2O-CCG against two representative baselines: classical CCG \cite{zeng2013solving} and Neur2RO \cite{dumouchelle2023deep}. For classical CCG, since grid search over $\mathcal{U} \subset \mathbb{R}^5$ is infeasible, the adversarial step uses multi-start random sampling ($50$ candidates) with exact QP evaluation. This serves as the gold-standard reference for solution quality. For Neur2RO, the adversarial subproblem becomes a MILP over the uncertainty set. For ellipsoidal and GMM sets, the MILP includes quadratic and big-M constraints respectively. 


\subsection{In-Distribution Results}

We evaluate all methods on the nominal problem instances where the training and test uncertainty sets are identical (in-distribution). For each uncertainty set type, we report: (i) Verified objective: the robust cost evaluated at each method's $u_0^*$ against the same multi-start adversarial oracle ($500$ candidates), ensuring fair comparison; (ii) computation time: total solve time including all CCG iterations; (iii) Number of scenarios: total scenarios generated before convergence; (iv) Optimality gap: relative gap to the CCG reference.

\begin{table}[tb]
\centering
\caption{In-distribution comparison across uncertainty set types.}
\label{tab:ind_results}
\begin{tabular}{@{}llrrr@{}}
\toprule
Set & Metric & CCG & Neur2RO & L2O-CCG \\
\midrule
\multirow{4}{*}{Box}
& Verified Obj.       & 25.60   & 25.64   & 25.66   \\
& Time (s)            & 87.24   & 10.99   & 0.78   \\
& \# Scenarios        & 2   & 3   & 1   \\
& Gap (\%)            & 0.0  & 0.1\%   & 0.2\%   \\
\midrule
\multirow{4}{*}{Polyhedral}
& Verified Obj.       & 25.60   & 25.63   & 24.64   \\
& Time (s)            & 83.86   & 81.28   & 1.37   \\
& \# Scenarios        & 2   & 3   & 1   \\
& Gap (\%)            & 0.0  & 0.1\%   & -3.7\%   \\
\midrule
\multirow{4}{*}{Ellipsoidal}
& Verified Obj.       & 22.88   & 21.19   & 21.46   \\
& Time (s)            & 753.53   & 59.32   & 0.74   \\
& \# Scenarios        & 21   & 2   & 1   \\
& Gap (\%)            & 0.0  &  7.4\%  &  6.2\%  \\
\midrule
\multirow{4}{*}{GMM}
& Verified Obj.       & 25.82   & 27.97   & 27.04   \\
& Time (s)            & 832.81   & 61.38   & 9.26   \\
& \# Scenarios        & 21   & 2   & 5   \\
& Gap (\%)            & 0.0  & 8.3\%   & 4.7\%   \\
\bottomrule
\end{tabular}
\end{table}

Table. \ref{tab:ind_results} compares the three methods under InD settings across four uncertainty set geometries. L2O-CCG consistently achieves the fastest solve times across all four geometries. The solution quality remains competitive: on box and polyhedral sets, the optimality gap is within 0.2\% and 3.7\%, respectively; on ellipsoidal and GMM sets, L2O-CCG actually achieves smaller gaps (6.2\% and 4.7\%) than Neur2RO (7.4\% and 8.3\%), suggesting that the learned optimizer navigates the non-convex adversarial landscape more effectively than the MILP reformulation in these geometries. The iteration number required by L2O-CCG is also notably low (1--5 scenarios), indicating that the L2O-CCG identifies high-quality worst-case scenarios from the first iteration.

\subsection{Out-of-Distribution Results}

A key advantage of our bi-level L2O-CCG is robustness to distribution shift. We train the L2O optimizer on a nominal uncertainty set and evaluate on perturbed sets at test time:
\begin{itemize}
    \item \textbf{Box}: Train with $(\theta, \Gamma) = (0.3, 1.0)$; test with $\theta \in \{0.2, 0.4, 0.5\}$ and $\Gamma \in \{0.8, 1.2, 1.5\}$.
    \item \textbf{Polyhedral}: Train with nominal ($\theta, \Gamma,H,h$); text with scaled $\theta,\Gamma$ and tightened/relaxed halfplance constraints $h' = s_h \cdot h$ for $s_h \in \{0.8, 1.2\}$.
    \item \textbf{Ellipsoid}: Train with nominal $\Sigma, \Gamma$; test with $\Gamma' = s \cdot \Gamma$ for $s \in \{0.7, 1.3, 1.5\}$ and rotated $\Sigma$.
    \item \textbf{GMM}: Train with nominal $\{(w_c, \mu_c, \Sigma_c)\}_{c=1}^3$; test with shifted component means $\mu_c' = \mu_c + \delta$ for $\delta \in \{0.05, 0.1\}$ and modified density threshold $\rho' \in \{0.5\rho, 2\rho\}$.
\end{itemize}

\begin{figure}[t]
    \centering
    \includegraphics[width=\columnwidth]{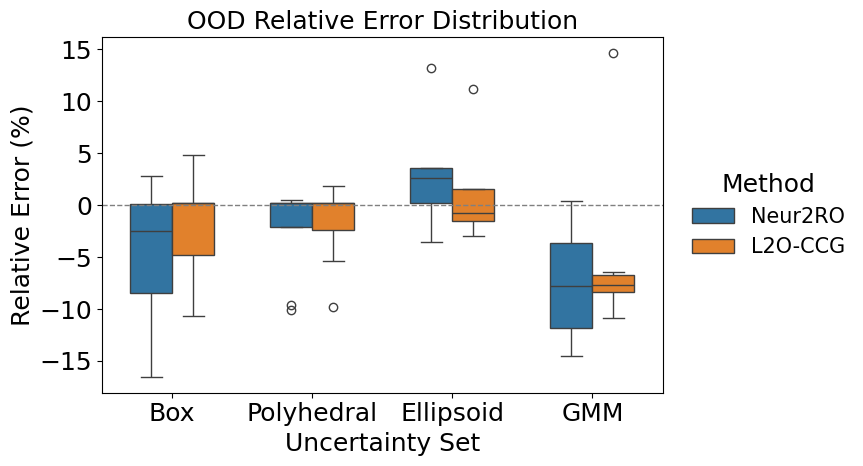}
    \caption{Relative error (compared with benchmark CCG) across out-of-distribution scenarios for different uncertainty sets. We evaluate four different uncertainty sets: box, polyhedral, elliposid, and GMM. L2O-CCG consistently achieves comparable or lower error than Neur2RO across all settings, while maintaining tighter distributions and fewer extreme deviations. The results indicate strong generalization performance of the proposed method under distributional shifts.}
    \label{fig:error}
\end{figure}
\begin{figure}[t]
    \centering
    \includegraphics[width=\columnwidth]{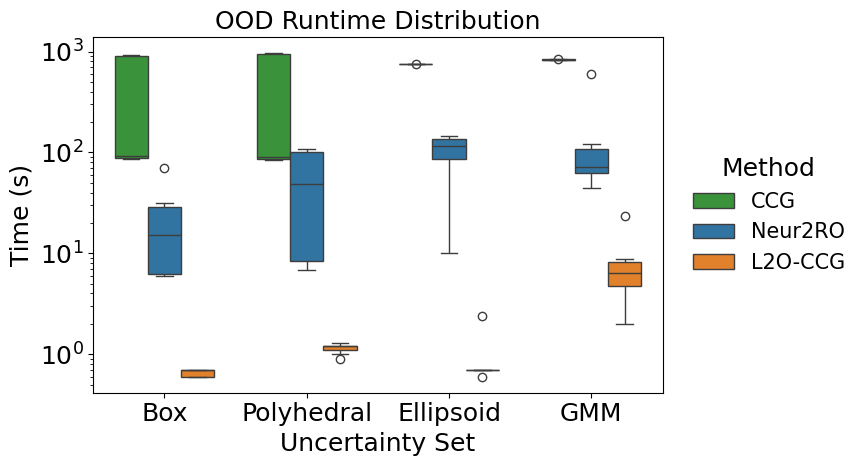}
    \caption{Computation time across out-of-distribution scenarios for different uncertainty sets. We evaluate four different uncertainty sets: box, polyhedral, elliposid, and GMM. L2O-CCG achieves orders-of-magnitude speedup compared to CCG and outperforms Neur2RO in most cases.}
    \label{fig:time}
\end{figure}

Fig.~\ref{fig:error} and Fig.~\ref{fig:time} compare generalization performance when the uncertainty set parameters shift from those used during training. Across all four geometries, L2O-CCG maintains smaller optimality gaps than Neur2RO. This is because since the LSTM never sees set parameters during training, its step-size policy is less affected by parameter shifts, and the proximal operator automatically adapts to the new geometry. Notably, the gap improvement of L2O-CCG over Neur2RO is more pronounced under OOD conditions than InD.


\section{Conclusions}

The adversarial subproblem in two-stage ARO is a key computational bottleneck that limits the scalability of decomposition-based algorithms. We proposes L2O-CCG, a modular bi-level framework that integrates learning-to-optimize into the CCG algorithm. By combining a NN surrogate of the recourse value function with a learned optimizer trained via algorithm unrolling, L2O-CCG eliminates the need for MILP-based adversarial solves while preserving the convergence guarantee of the CCG algorithm through exact verification. The set-agnostic design, where uncertainty set parameters enter only through proximal operators and are decoupled from the learned optimizer parameters, enables a single trained model to generalize across different geometry without retraining. We establish OOD convergence bounds showing how the optimizer's trajectory deviation is controlled by the distribution shift. Experiments on a robust MPC benchmark confirm that L2O-CCG achieves significant speedups over classical CCG and MILP-based learning approaches (Neur2RO) with comparable solution quality across four uncertainty geometries, under both in-distribution and out-of-distribution cases.


\bibliographystyle{ieeetr}
\bibliography{references.bib}

\end{document}